\documentclass{article}
\usepackage[a4paper, total={6in, 8in}]{geometry}
\usepackage{amssymb}
\usepackage{graphicx,epstopdf}
\usepackage{gensymb}
\usepackage{epsfig}
\usepackage{color}
\usepackage{epstopdf}
\usepackage{amsmath}
\usepackage{caption}
\usepackage{subcaption}
\usepackage{authblk}
\usepackage[super,comma]{natbib}
\usepackage{url}

\newcommand{\grl}{    {Geophys. Res. Lett.}}
\newcommand{\jgr}{    {J. Geophys. Res.}}
\newcommand{\ssr}{    {Space Sci. Rev.}}
\newcommand{\planss}{    {Plan. Sp. Sci.}}

\begin{document}

%\preprint{}

\title{Charged particle scattering in dipolarized magnetotail} %Title of paper

\author[1,2]{A. S. Lukin}
\author[1,3]{A. V. Artemyev}
\author[1]{A. A. Petrukovich}
\author[3]{X.-J. Zhang}
\affil[1]{Space Research Institute of the Russian Academy of Sciences (IKI), 84/32 Profsoyuznaya Str, Moscow, Russia, 117997; as.lukin.phys@gmail.com}
\affil[2]{Faculty of Physics, National Research University Higher School of Economics, 21/4 Staraya Basmannaya Ulitsa, Moscow, Russia, 105066}
\affil[3]{Department of Earth, Planetary, and Space Sciences, University of California, 595 Charles E Young Dr E, Los Angeles, CA, California, USA, 90095; aartemyev@igpp.ucla.edu}

%Provide a complete mailing address for each author. Please include the Department, Company/University, City, State, Postal code, and Country for each author of the submitted manuscript. The byline must be placed after the title and before the abstract. Please add the postal codes to the first four affiliations currently listed in your paper.

%\date{\today}

\maketitle

\begin{abstract}
The Earth's magnetotail is characterized by stretched magnetic field lines. Energetic particles are effectively scattered due to the field-line curvature, which then leads to isotropization of energetic particle distributions and particle precipitation to the Earth's atmosphere. Measurements of these precipitation at low-altitude spacecraft are thus often used to remotely probe the magnetotail current sheet configuration. This configuration may include spatially localized maxima of the curvature radius at equator (due to localized humps of the equatorial magnetic field magnitude) that reduce the energetic particle scattering and precipitation. Therefore, the measured precipitation patterns are related to the spatial distribution of the equatorial curvature radius that is determined by the magnetotail current sheet configuration. In this study, we show that, contrary to previous thoughts, the magnetic field line configuration with the localized curvature radius maximum can actually enhance the scattering and subsequent precipitation. The spatially localized magnetic field dipolarization (magnetic field humps) can significantly curve magnetic field lines far from the equator and create off-equatorial minima in the curvature radius. Scattering of energetic particles in these off-equatorial regions alters the scattering (and precipitation) patterns, which has not been studied yet. We discuss our results in the context of remote-sensing the magnetotail current sheet configuration with low-altitude spacecraft measurements.
\end{abstract}

\maketitle
%\maketitle must follow title, authors, abstract and \pacs

\section*{Introduction}
Configuration of the Earth’s magnetotail current sheet determines its stability and controls many energetic events, e.g., magnetospheric substorms \cite{Baker96,Angelopoulos08, Angelopoulos20} and plasma convection \cite{Sergeev96, Milan19}. In-situ spacecraft measurements of the magnetotail plasma and magnetic field are rather limited, as spacecraft (even multiple) can only simultaneously probe few spatially localized regions (see Refs. \cite{Runov06,Runov09,Sergeev06,Petrukovich15:ssr,Artemyev21:grl} for examples of probing of the magnetotail current sheet configuration with in-situ multi-spacecraft missions). Thus, alternative methods to examine the current sheet configuration can be especially useful. Statistical methods \cite{Tsyganenko13} have then been used to reconstruct the current sheet configuration by fitting to the model \cite{Tsyganenko95, Tsyganenko02, Tsyganenko&Sitnov05} or via data mining approaches \cite{Sitnov19:jgr, Stephens16,Stephens19}. This set of methods provides quite accurate description of the magnetotail current sheet, but very restricted to the time-scale of evolution of the current sheet configuration, i.e. the statistical reconstruction shows some averaged (typical) configuration that may slowly evolve with the time-scale of the evolution of input model parameters (that are often geomagnetic indexes, see Refs. \cite{Tsyganenko13, Sitnov21}). The second group of methods considers low altitude spacecraft measurements of charged particle precipitations from the magnetotail for reconstruction of instantaneous current sheet configuration \cite{Sergeev11, Sergeev12:IB, Dubyagin13}. These methods are essentially based on the theory of charged particle scattering in the curved magnetic field lines of the magnetotail current sheet \cite{Imhof79, Sergeev&Tsyganenko82, Sergeev83, BZ89, Delcourt95}, and our study develops this theory for a complex current sheet configurations.

\begin{figure*}
\centering
\includegraphics[width=0.95\textwidth]{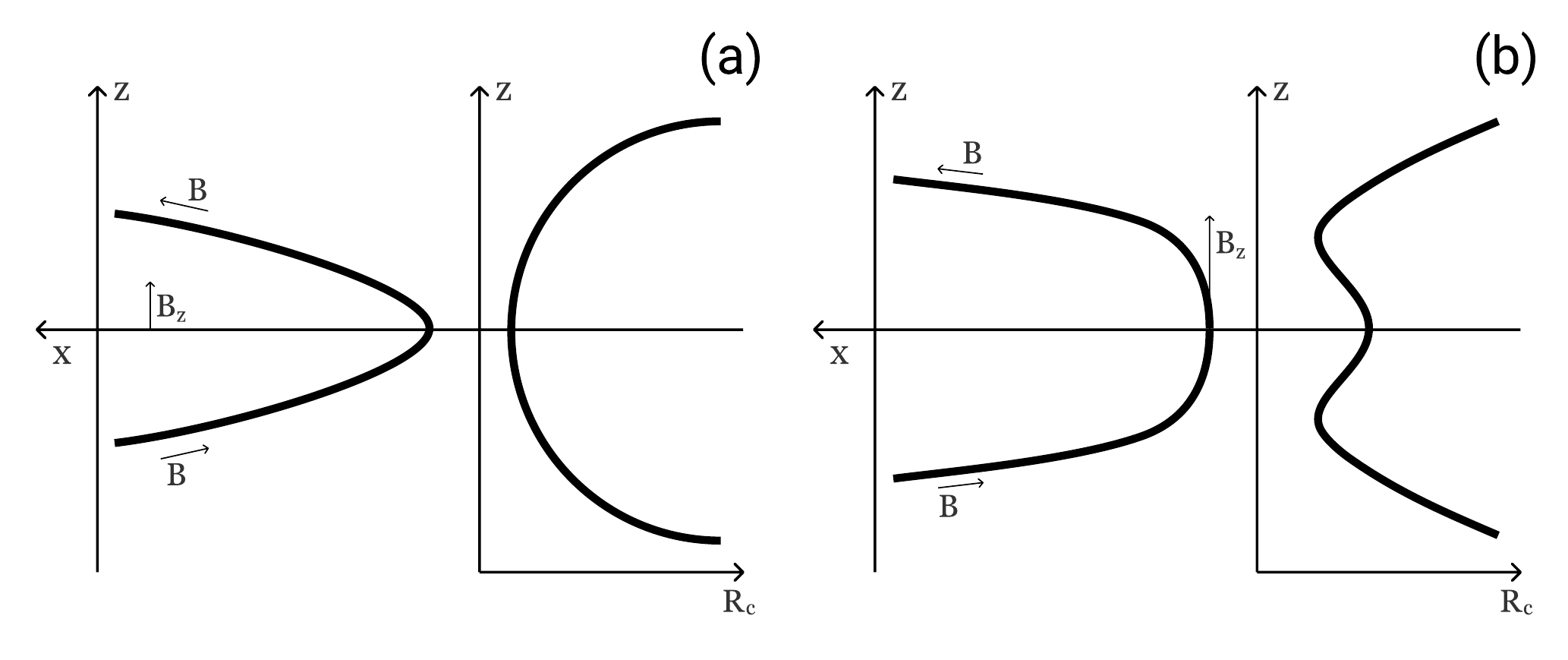}
\caption{Schematic view of magnetic field lines and field line curvature radius in the magnetotail current sheet for two typical configurations: (a) quite-time thin current sheet, (b) current sheet with the embedded dipolarization front.}
\label{fig1}
\end{figure*}

Schematic in Fig. \ref{fig1}(a) shows that magnetic field lines in a quite-time thin current sheet are well stretched due to smallness of the ratio of magnetic field components, $B_z/B_{0x}\ll 1$ (see for details Refs. \cite{Runov06,Petrukovich15:ssr}). Magnetized electrons and ions are scattered in such magnetic field configuration, i.e. there is a jump of magnetic moment $\mu=mv_\perp^2/B$ ($B=|{\bf B}|$ and $v_\perp$ is the transverse component of particle velocity) when particles cross the region with the smallest $R_c/\rho$ ($R_c$ is magnetic field line curvature radius, $\rho$ is the particle gyroradious). For the most typical $R_c/\rho>1$ magnetotail configuration \cite{Runov06,Artemyev16:jgr:pressure} such jumps of a magnetic moment can be considered within the theory of the adiabatic invariant destruction in the slow-fast nonlinear systems \cite{bookLL:mech60,Slutskin64, Neishtadt00}. The basic model of this destruction has been constructed for magnetic traps in laboratory plasma \cite{Chirikov78,Chirikov79,Howard71,Cohen78} and then generalized to the current sheets in the planetary magnetotails \cite{Birmingham84,BZ89}. Comprehensive numerical investigations result in the empirical models of $\Delta\mu$ jump as a function of $R_c/\rho$. For a fixed particle energy these models provide $\Delta\mu$ as a function of magnetic field configuration \cite{Delcourt94:scattering,Delcourt96:precipitation, Young02, Young08}. The main model prediction is threshold of $B_z/B_{0x}$ for strong scattering (significant $\Delta\mu$). Such threshold provides the basement for analysis of low-altitude observations for remote probing the magnetotail current sheet configuration \cite{Yahnin97, Sergeev15, Dubyagin18}. These models, however, have been developed for the {\it classical} current sheet with a single $R_c$ minimum at the equatorial ($B_x=0$) plane (see Fig. \ref{fig1}(a)).

If there is only equatorial $R_c$ minimum, any suppression of equatorial scattering (decrease of precipitating fluxes on the low-altitude spacecraft) should be interpreted as a local increase of $B_z/B_0$ ratio\cite{Sergeev18:grl, Dubyagin21}. This interpretation may be invalid for more complex current sheet configurations with off-equatorial $R_c$ minima \cite{Delcourt06,Delcourt06:AdSR}. One of such configurations is shown in Fig. \ref{fig1}(b) where so-called dipolarized current sheet is presented. The transient magnetic reconnection in the middle (far from the Earth) magnetotail generates very large amplitude perturbations of $B_z$ (dipolarization fronts, see Refs. \cite{Sitnov09, Runov09grl,Angelopoulos13,Fu13:grl}). These perturbations propagate earthward and breaking in the near-Earth current sheet \cite{Dubyagin11, Liu14:DF}. Thus, in the near-Earth current sheet we can observe a magnetic field configuration with a sharp $B_z$ gradient ($\partial B_z/\partial x$) and strong $B_z$ enhancement. Such an enhancement should decrease the euqatorial $R_c$ and suppress charged particle scattering, but large $\partial B_z/\partial x$ may create additional off-equatorial $R_c$ minima and enhance the scattering. Therefore, for accurate interpretation of precipitating particle fluxes on low-altitude spacecraft there is a need for parametric investigation of the charged particle scattering in the magnetotail current sheet configuration with the embedded dipolarization front.

In this study we investigate charged particle scattering in such magnetotail current sheet configuration. First, we introduce the magnetic field model in Sect. ~\ref{sec:model}. Then the massive test particle simulations are used to show how the current sheet and dipolarization front parameters affect the efficiency of charged particle scattering. Main results of these simulations are shown in Sect. ~\ref{sec:simulations}. We also modify the analytical model of charged particle scattering in the current sheet \cite{Birmingham84} to explain effects of off-equatorial $R_c$ minima (see Sect. ~\ref{sec:theory}). Then we discuss obtained results and list main conclusions in Sect. ~\ref{sec:conclusions}.

\section{Charged particle dynamics \label{sec:model}}
The magnetic field configuration of the magnetotail current sheet around the equatorial plane (where main particle scattering occurs) can be fitted by a simple 1D model with $B_z=const$ and $B_x=B_{0x}\cdot(z/L_z)$, where $L_z$ is the current sheet thickness. The curvature radius for this model has a minimum $\min R_c= L_{z}B_{z}/B_{0x}$ at the equator $z=0$, where the charged particle gyroradius maximizes $\rho=\sqrt{2Emc^2}/eB_{z}$ where $E$, $m$, $e$ are particle energy, mass, and charge. To describe the dipolarization front embedded into the current sheet, we modify $B_z$ field as:
\begin{equation}
\label{eq:field}
    B_{z}=B_{+z} + 0.5\left(B_{-z}-B_{+z}\right)\left(1-\tanh\left(x/L_x\right)\right)
\end{equation}
where $L_x$ is the front thickness, and we locate the dipolarization front at $x=0$. Far from the dipolarization front $B_z=B_{-z}$ for $x<0$ and $B_z=B_{+z}$ for $x>0$ where $B_{-z}>B_{+z}$.  Three main system parameters are $B_{+z}/B_{0x}$, $B_{-z}/B_{0x}$, and $L_x/L_z$. Let us discuss ranges of these parameters derived from spacecraft observations in the near-Earth magnetotail. The current sheet thickness $L_z$ varies from $500$ km to $5000$ km for most of observed magnetotail current sheets \cite{Petrukovich15:ssr,Artemyev16:jgr:pressure}, and $L_z$ is larger for post-dipolarization current sheet\cite{Yushkov21}. The dipolarization front thickness $L_x$ is generally smaller than $1000$ km (see Refs. \cite{Runov11pss, Hwang11}) and can be as small as $300-500$ km (see e.g., Ref. \cite{Khotyaintsev11, Fu12:dipolarization:structure}). The magnetic field magnitude of the dipolarization front is comparable to the current sheet magnetic field, $B_{-z}\sim B_{0x}$, and is larger than the background $B_{+z}\sim B_{-z}\times[0.1,0.5]$ (see statistics in Refs. \cite{Runov11jgr, Runov15}).

\begin{figure}
\centering
\includegraphics[width=0.5\textwidth]{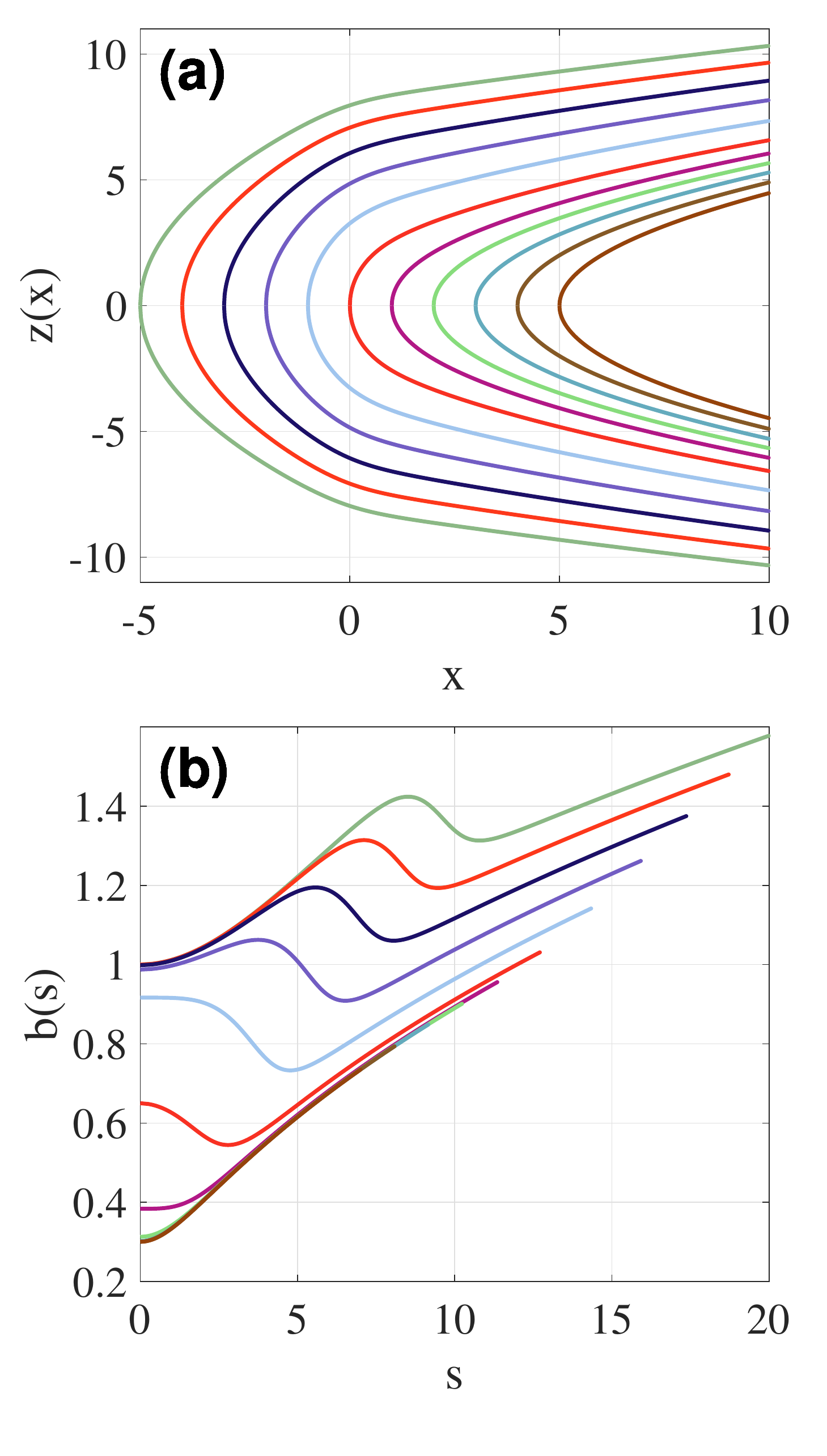}
\caption{Magnetic field lines in the magnetotail current sheet with the embedded dipolarization front: $B_x=B_{0x}\cdot(z/L_z)$, $B_{z}$ given by Eq. (\ref{eq:field}), $B_{-z}/B_{0x}=1$, $B_{+z}/B_{0x}=0.3$, $L_x/L_z=0.15$.}
\label{fig2}
\end{figure}

Figure \ref{fig2}(a) shows magnetic field line configuration for typical parameters of the current sheet and dipolarization front. Ahead the front ($x_0>0$) there are stretched magnetic field lines (parabolic shape $x\sim z^2$) with a curvature radius minimum, $R_c=L_zB_{z}/B_{0x}$, at the equatorial plane, where energetic particles are expected to be scattered. At the front the magnetic field line configuration changes and the curvature radius at the equator increases significantly. For some field lines there is almost no curvature at the equator, $R_c\to\infty$. However, due to the strong $dB_z/dx\sim1/L_x$ gradient, there are two new off-equatorial $R_c$ minima. Behind the dipolarization front ($x_0<0$) the current sheet configuration includes parabolic magnetic field lines around the equator (but the curvature radius there $R_c=L_zB_{-z}/B_{0x}$ is much larger than this radius ahead the front). These lines cross the front $dB_z/dx$ gradient, where a local minimum of the curvature radius may occur. Thus, energetic particle scattering in such magnetic field configuration strongly depends on the particle pitch-angle (i.e., location of particle mirror points where $E=\mu B$), because particles with smaller pitch-angles can reach off-equatorial minima of the curvature radius.

The scattering efficiency is determined by $\kappa=\min\sqrt{R_c/\rho}$ parameter \cite{Zelenyi13:UFN}, and thus depends on the magnetic field magnitude profile $B(s)$ along field lines: $s=\int_0^z{B(x,z)dz/B_z(x)}$ where $dx/dz=B_x(z)/B_z(x)$ is the magnetic field line equation. Contrast to the classical current sheet configuration with $B_z=const$, in the configuration from Fig. \ref{fig2} the profile $B(s)$ varies with the $x_0$, the field-line root position at the equator. Figure \ref{fig2}(b) shows several $B(s)$ profiles ahead the front, at the front, and behind the front (color coding are the same as in panel (a)). There are clear off-equatorial $B$ minima for $x_0$ around the front.

\section{Numerical results \label{sec:simulations}}
To investigate charged particle scattering in the current sheet with the embedded dipolarization front we numerically integrate multiple trajectories described by 2D equations of motion:
$ {\bf \dot r} = {\bf v}$, ${\bf \dot v} =  - e\left[ {{\bf v} \times {\bf B}} \right]/m_e c$, where we use electron mass and charge (note the obtained results are applicable to ion scattering as well with the proper renormalization of energy to keep the same $\kappa$).

For $50$ values of $x_0\in[-5,5]$ values and $12$ equatorial pitch-angle values we numerically integrate $50\times12\times10^{4}=6\times10^6$ orbits. Each orbit starts at the mirror point ${\bf v}\cdot{\bf B}=0$ above the equator (i.e. at $z>0$) and ends on the opposite side of the equator at another mirror point. Initial $\mu_0=mv_{\perp}^2/B=mv^2/B$ and final  $\mu_f$ magnetic moments are calculated at mirror points, and this procedure reduces the magnetic moment fluctuations \cite{Anderson97, Eshetu18}. These fluctuations are due to the fact that $\mu$ is the adiabatic (approximate) invariant and $mv_{\perp}^2/B$ is the leading order approximation, whereas more accurate equations for $\mu$ can be derived using the improving procedure for adiabatic invariants\cite{bookAKN06}.

\begin{figure*}
\centering
\includegraphics[width=0.95\textwidth]{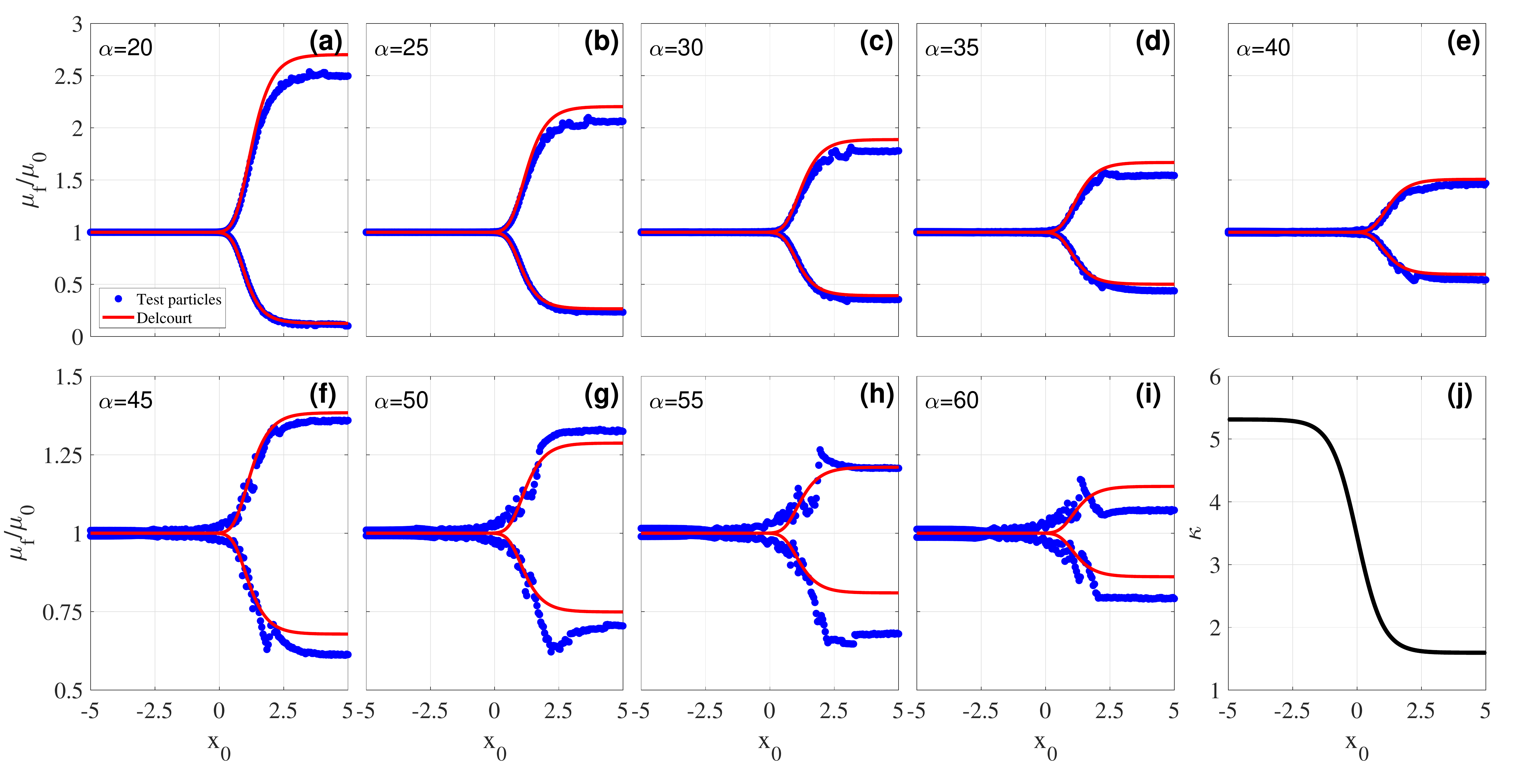}
\caption{Comparison of $\mu_{f}/\mu_{0}$ for the analytical model of charged particle scattering in the current sheet\cite{Delcourt96:precipitation}(shown by red curves) and numerical results (shown by blue circles). Panels (a)-(i) correspond to different equatorial particle pitch angles (see the upper left corner of each panel). System parameters are: $L_x/L_z=10$, $B_{-z}/B_{0x}=1$, $B_{+z}/B_{0x}=0.3$. Particle energy is chosen in such way to have $\kappa\sim 1.5$ for large $x_0>0$. Profile of $\kappa(x_0)$ evaluated at the equator is shown in the panel (j).}
\label{fig3}
\end{figure*}

Figure \ref{fig3} shows boundaries $\mu_f/\mu_0$ of the distribution for different $x_0$ and $L_x/L_z=10$ (very smoothed dipolarization front). We show $\mu_f/\mu_0$ for 5\% of particles with largest $\mu_f/\mu_0$ and 5\% of particles with the smallest $\mu_f/\mu_0$ (i.e., all $\mu_f/\mu_0$ are distributed somewhere between these two boundaries). This form of presentation allows us to compare numerical simulation results with analytical models of $\mu_f/\mu_0$ derived for the current sheet configuration (see Refs. \cite{Delcourt96:precipitation,Shustov15}; note these two models provide almost identical $\mu_f/\mu_0$ for equatorial pitch-angles below $\sim 60^\circ$ that are most interesting in the context of investigation of energetic particle scattering to the low altitudes). To make such a comparison we calculate $\kappa=\min\sqrt{R_c/\rho}$ for each $x_0$ and then evaluate the model $\mu_f/\mu_0$ for this $\kappa$. Note the model, derived in \cite{Delcourt96:precipitation} has been constructed for the current sheet configuration with $\kappa$ determined by the equatorial magnetic field line curvature, and thus model/simulations difference are due to off-equatorial scattering effects. Figure \ref{fig3} shows that at large $x_0$, where Eq. (\ref{eq:field}) gives $B_z=B_{+z}$, our numerical results are very close to the model, what verifies our scheme of $\mu_f/\mu_0$ calculations. Moreover, due to weak the $dB_z/dx$ gradient for this simulation run ($L_x/L_z=10$), even at $x_0$ around zero (the dipolariaztion front location, field-aligned particle scattering is well described by the model developed for the current sheet configuration. The model/simulations difference can be seen only for intermediate pitch-angles.

\begin{figure*}
\centering
\includegraphics[width=0.95\textwidth]{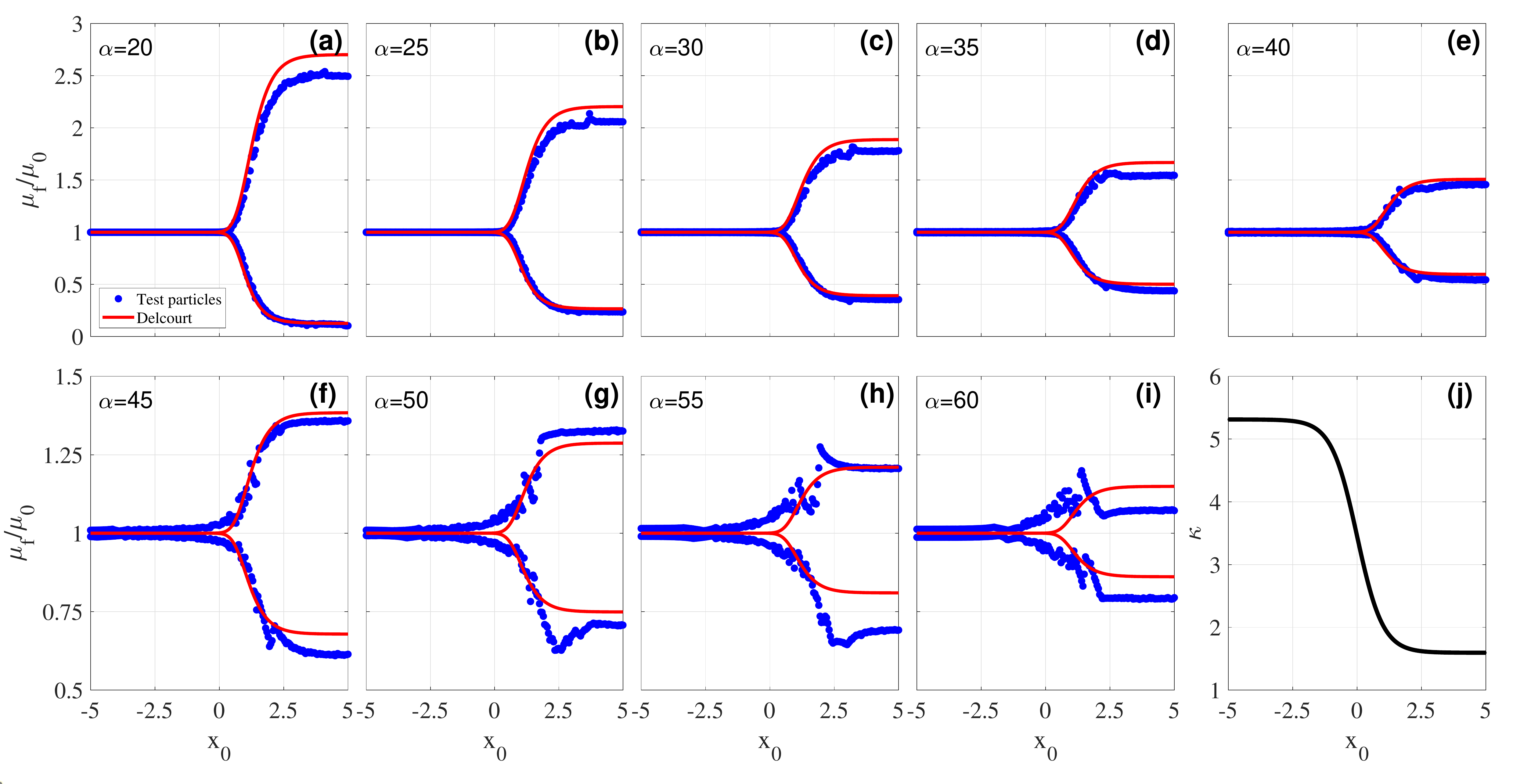}
\caption{The same as in Fig. \ref{fig3}, but for $L_x/L_z=1$.}
\label{fig4}
\end{figure*}

Figure \ref{fig4} shows a scattering efficiency for moderately thin dipolarization front with $L_x/L_z=1$. Small pitch-angle particles are scattered in agreement with predictions of the model constructed for the current sheet configuration, i.e. $dB_z/dx$ is not sufficiently strong to change scattering. For intermediate pitch-angels the effect of $dB_z/dx$ is seen better: there is a clear peak of scattering efficiency (maximum of $\mu_f/\mu_0$) around $x_0\sim 0$ for $\alpha>50^\circ$. Therefore, the dipolarization front indeed can change the pattern of scattering and makes scattering spatially non-monotonic. But for $L_x/L_z=1$ this effect is seen only for intermediate pitch-angles, far from the pitch-angle range corresponding to particle scattering to low-altitudes.

\begin{figure*}
\centering
\includegraphics[width=0.95\textwidth]{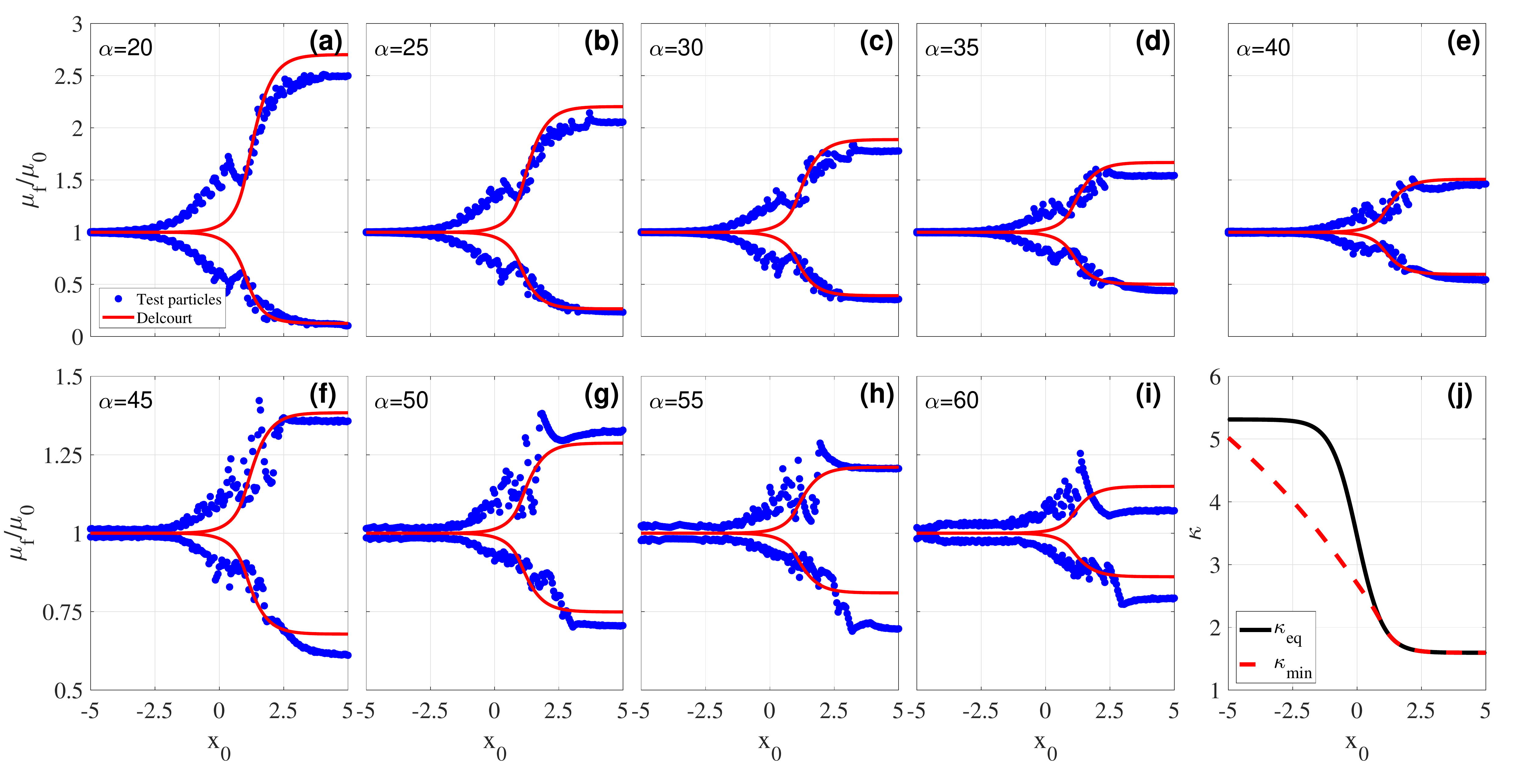}
\caption{The same as in Fig. \ref{fig3}, but for $L_x/L_z=0.15$. Red dashed line in panel (j) shows the minimal $\kappa$ along field line, while solid black line shows equatorial $\kappa$.}
\label{fig5}
\end{figure*}

Figure \ref{fig5} shows results for the thin dipolarization front with $L_x/L_z=0.15$. So strong $dB_z/dx$ moves the minimum $\kappa$ position away from the equator (see Fig. \ref{fig2}(b)) and creates local (in $x_0$) maximum of the scattering efficiency ($\mu_f/\mu_0$ peaks) at $x_0\sim 0$ for a wide pitch-angle range, including small pitch-angles. Such peaks of $\mu_f/\mu_0$ (or local minima of $\mu_f/\mu_0$) would be seen by low-altitude spacecraft as transient decreases/increases of precipitations along the spacecraft orbit \cite{Sergeev18:grl}. Note $\mu_f/\mu_0$ peaks/minima are not due to $\kappa(x_0)$ non-monotonical profile (see Fig. \ref{fig5}(f) showing equatorial $\kappa(x_0)$ and real $\kappa(x_0)=\min\sqrt{R_c(s)/\rho(s)}$ profiles), but due to change of the scattering pattern: instead of the single scattering at the equator, typical for the current sheet configuration, the strong $dB_z/dx$ gradient creates two off-equatorial locations of scattering. Such scattering cannot be characterized by the single $\kappa$ parameter and the scattering model should take into the account actual $B$ profile along magnetic field lines.

\begin{figure}
\centering
\includegraphics[width=0.5\textwidth]{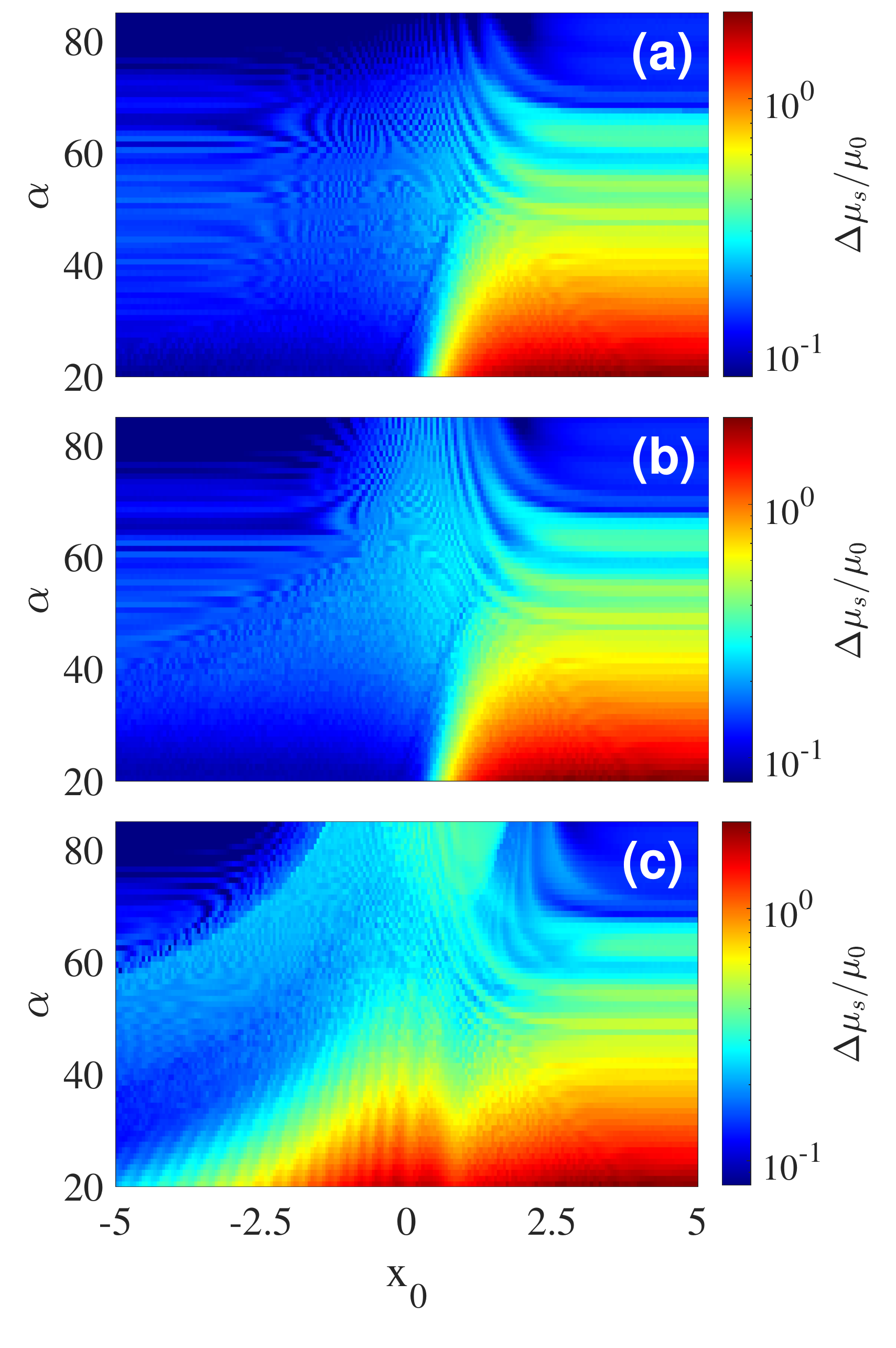}
\caption{Distribution of $\Delta\mu_s/\mu_{0}$ in $(x_0, \alpha)$ space for system parameters $B_{-z}/B_{0x}=1$, $B_{+z}/B_{0x}=0.3$ and (a) $L_x/L_z=10$, (b) $L_x/L_z=1$, (c) $L_x/L_z=0.15$.}
\label{fig6}
\end{figure}

Figure \ref{fig6} compares patters of scattering $\Delta\mu_s/\mu_0=\sqrt{\langle\left(\mu_f-\mu_0\right)^2\rangle}/\mu_0$, where $\langle ... \rangle$ denotes the ensemble averaging, for three $L_x/L_z$ values. The weak $dB_z/dx$ gradient separates two regions with weak ($x_0<0$) and strong ($x_0>0$) scattering of all pitch-angles for $L_x/L_z=10$ (see Fig. \ref{fig6}(a)). The stronger $dB_z/dx$ gradient creates a local peak of scattering for intermediate pitch-angles around $x_0\sim0$, but does not affect weak scattering of small pitch-angle particles (see Fig. \ref{fig6}(a)). The strong $dB_z/dx$ gradient creates the secondary maximum of scattering around $x_0\sim 0$ for all pitch-angles.

\begin{figure}
\centering
\includegraphics[width=0.5\textwidth]{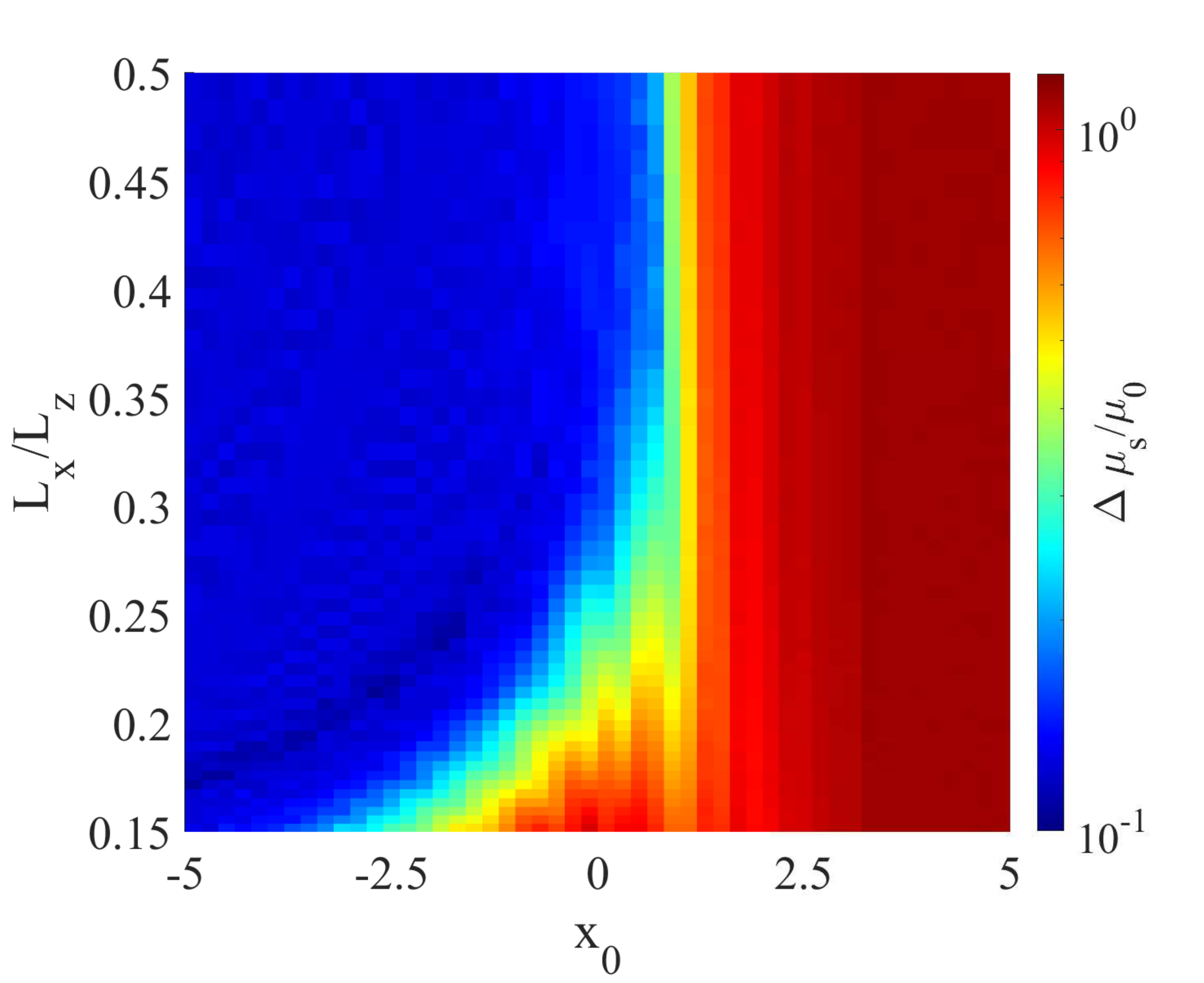}
\caption{Distribution of $\Delta\mu/\mu_{0}$ in $(x_0, L_x/L_z)$ space for $\alpha=30^\circ$.}
\label{fig7}
\end{figure}

Figure \ref{fig7} shows the parametric investigation of the scattering efficiency. We fix $\alpha=30^\circ$ and consider various $L_x/L_z\in[0.15,0.5]$. The second peak of the scattering around the dipolarization front $x_0\sim 0$ occurs for $L_x/L_z\sim 0.2-0.25$ for $B_{\pm z}/B_{0x}=0.3,1$ magnetic field magnitudes. Such $L_x/L_z$ values are quite typical for the dipolarized mangnetotail current sheet with $L_z\sim 3000$km and $L_x\sim 300-1000$ km (see Refs. \cite{Runov11pss,Runov12}). To confirm this double-peak scattering pattern we consider analytical theory of scattering in the next section.

\section{Analytical estimates \label{sec:theory}}
The magnetic moment is the adiabatic invariant that should conserve with the exponential accuracy for the charged particle motion in a slowly time varying magnetic field (or a magnetic field slowly depending on coordinates) \citep{Hertweck&Schlueter57, Dykhne60, Vandervoort61}. The general approach of evaluation of $\mu$ change has been proposed in Ref. \cite{Slutskin64}, but we are interested here only in the main part of this change, namely exponential term $\Delta\mu/\mu_0\sim \exp(-f(\alpha)\kappa^2)$. For slow-fast dynamical systems (e.g., when gyrorotation is much faster than bounce motion) the general approach of estimate of this exponential term can be found in Refs. \cite{Neishtadt00, Su12:RCD}. Such time separation of gyrorotation and bounce motion allows introduction of the magnetic field along the orbit as a function of a field-aligned coordinate, $B(s)$. Then, equation for function $f$ takes the form \cite{Birmingham84}:
\begin{eqnarray}
f(\mu ^* ) = \int_0^{\ell^*}\frac{b(\ell)d\ell}{\sqrt{1-\mu^* b(\ell)}}
\label{eq:f}
\end{eqnarray}
where $\mu^*=\sin^2\alpha$ is the normalized magnetic moment independent on energy, $\alpha$ is the equatorial pitch-angle, $\ell=\Im(s)$ is the imaginary part of the dimensionless coordinate along magnetic field line, $\ell^*$ is a solution of $b(\Re(s)+i\ell)=0$ equation, and $b(s)=B(s)/B_{0x}$ is dimensionless magnetic field magnitude. Being calculated, this exponential factor $\sim  f(\mu^*)\kappa^2$ can explain the principal dependence of the scattering efficiency ($\Delta\mu/\mu_0$) on $\kappa$ and equatorial pitch-angle. Note we restrict our consideration to this factor only and do not calculate the pre-exponential multiplication coefficient in $\Delta\mu/\mu_0$ expression. In Eq. (\ref{eq:f}) the integration is performed along the trajectory with $\mu^*=const$, but $\mu^*$ is only an approximate integral of motion that oscillates with an amplitude $\sim \kappa^{-2}$. Such a difference between actual and $\mu^*=const$ trajectories can affect the accuracy of $\Delta\mu$ evaluation \cite{Dykhne&Chaplik61}. This effect does not change the exponential factor, but can change the pre-exponential multiplication coefficient \cite{Neishtadt84, Neishtadt00}. Therefore, test particle approaches are generally used to estimate this coefficient \cite{Anderson97, Young02}, and we do not derive it theoretically.

To describe $b(s)$ profiles from Fig. \ref{fig2}(b) in simpler form allowing calculation $b(\Re(s)+i\ell)=0$ roots, we use the analytical model:
\begin{equation}
b_m  = b_{eq}  + as^2  - d\exp \left( { - \frac{{\left( {s - s_{\min } } \right)^2 }}{{\delta s^2 }}} \right) \label{modelField}
\end{equation}
%\begin{equation}\label{modelField}
%\begin{cases}
%b_m(s)=b_{eq}+c_2s^2-c_{\exp} \exp{-\frac{\left(s-s_{\min}\right)^2}{\Delta}} \; \text{if} \; \exists s_{\min}>0 \\
%b_m(s)=b_{eq}+c_2s^2 \; \text{if} \; \nexists s_{\min}>0
%\end{cases}
%\end{equation}
where $a=5/1000$, $\delta s^2=0.75$, $b_{eq}=b_{eq}(x_0)$ is the normalized equatorial magnetic field, $d=b_{eq}(x_0)+c_2s_{\min}^2-b(s_{\min})$ for $s_{\min}>0$, and $s_{min}$ is the coordinate of $b(\Re(s))$ minimum from  Section \ref{sec:model}.

\begin{figure}
\centering
\includegraphics[width=0.5\textwidth]{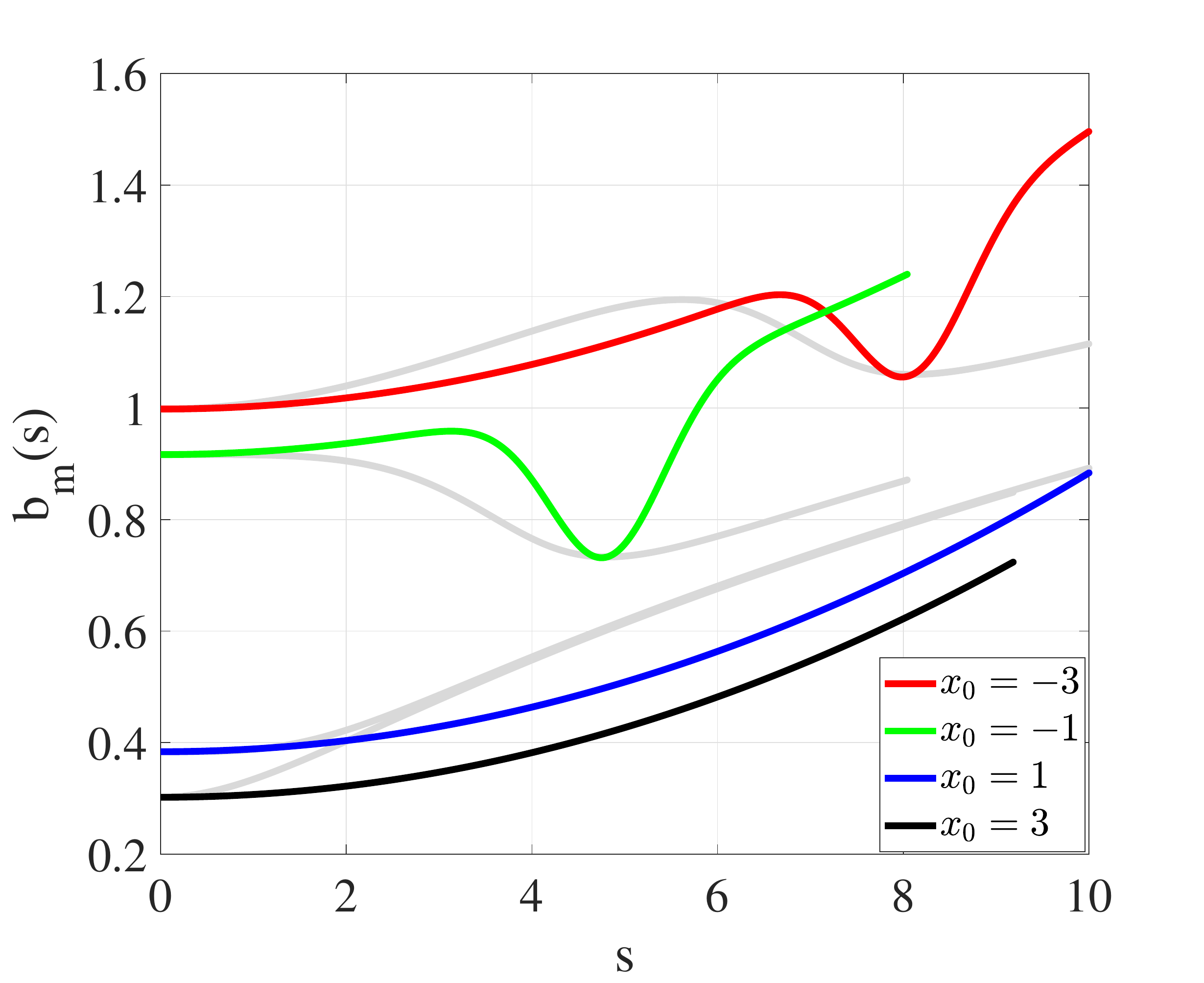}
\caption{Model magnetic field profiles $b_m(s)$ that mimic results from Fig. \ref{fig3}.}
\label{fig8}
\end{figure}

Model (\ref{eq:field}) is much simpler than the full field model from Section \ref{sec:model} and provides almost analytical solutions for $b(\Re(s)+i\ell)=0$. Moreover, this model is characterised by a finite $d^3b/ds^3|_{s=0}$ for all $x_0$ and do not have $d^3b/ds^3|_{s=0}\sim 0$ (see Fig. \ref{fig8}). Presence of a zero $d^3b/ds^3|_{s=0}$ (i.e. field line flattering at the equator) makes the standard scattering model\cite{Birmingham84} inapplicable, and requires derivation of more complecated equations instead Eq. (\ref{eq:f}), see, e.g., discussion in Ref. \cite{Vasiliev12}.

Figure \ref{fig8} shows that model (\ref{modelField}) reproduces well the main features of the magnetic field $B(s)$ profiles for different $x_0$. For large $x_0$ model shows the parabolic $B(s)$ with a single minimum at $s=0$, and around $x_0\sim 0$ model shows $B(s)$ profiles with off-equatorial minima.

%\todo{Comparison of Figs. \ref{fig3} and \ref{fig3} shows that analytical model for current sheet scattering describes well the scattering effect for a single-minimum $B(s)$, whereas multi-minimum $B(s)$ and $B(s)$ with the flatten profiles can result in more effective scattering.}

\begin{figure}
\centering
\includegraphics[width=0.5\textwidth]{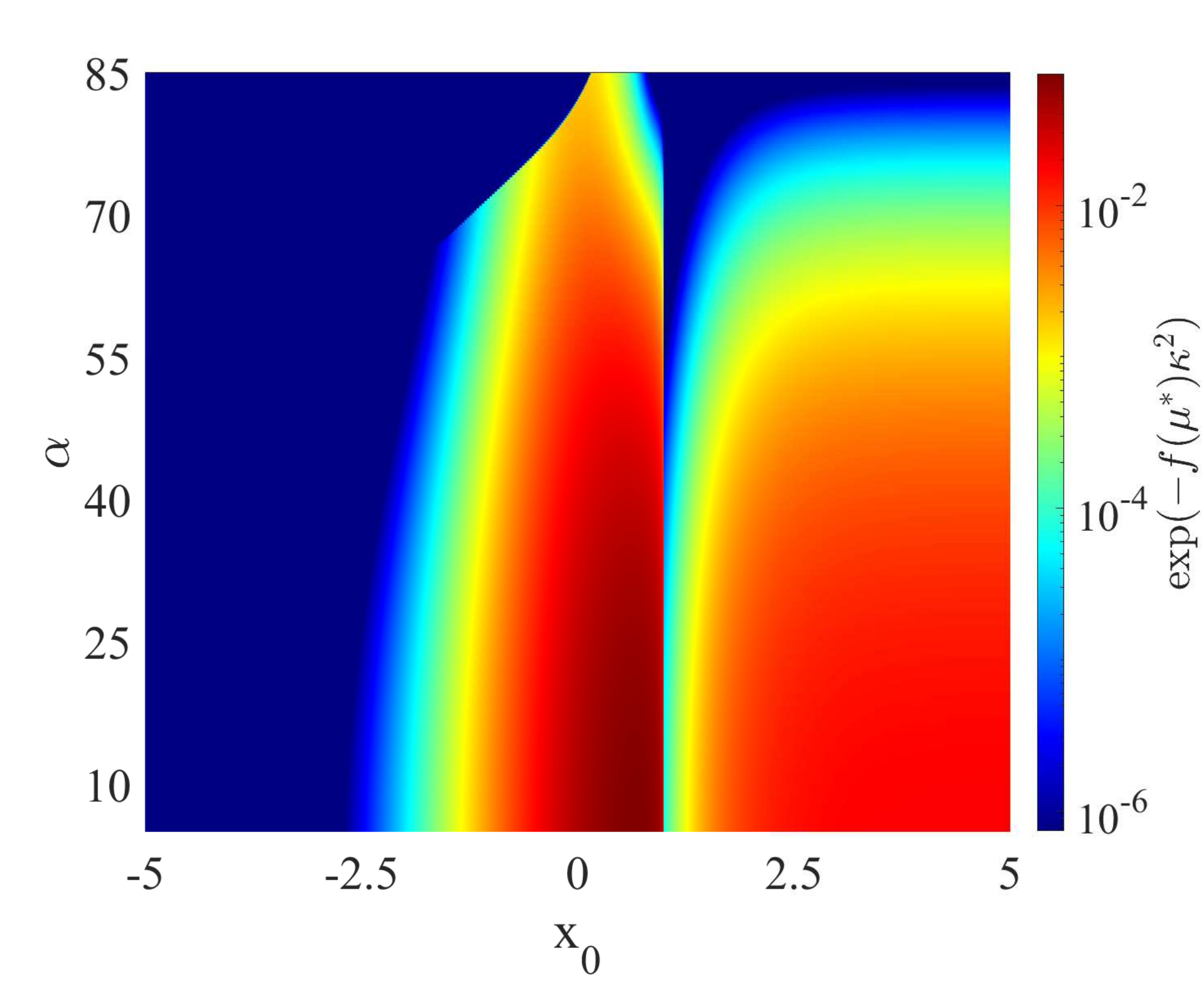}
\caption{Distribution of the theoretical {\it scattering} factor in $(\alpha,x_0)$ space for $b_m(s)$ model from Fig. \ref{fig8} and $\kappa_{\min}$ profile shown in Fig. \ref{fig5}(j).}
\label{fig9}
\end{figure}

Using model magnetic field (\ref{modelField}) and $\kappa(x_0)$ from Fig. \ref{fig5}(j), we plot exponential factor $\kappa^2f(\mu^*)$ in Fig. \ref{fig9}. Although this calculation does not take into account the pre-exponential multiplication factor, the main feature (two peak scattering, at $x_0>2$ and $x_0\sim 0$) is well seen (compare Figs. \ref{fig9} and \ref{fig5}). The second peak of scattering in the region of large $b_{eq}$ (large $B_z$) is due to off-equatorial magnetic field minima (and corresponding curvature maxima). This effect can be found only in the current sheet with embedded dipolarization front having a sharp gradient $dB_z/dx$. Analytical estimates on Fig. \ref{fig9} confirm the main numerical results of off-equatorial particle scattering at dipolarization front.

\section{Discussion and Conclusions \label{sec:conclusions}}
There is the quite powerful approach of the magnetotail current sheet probing with precipitating energetic particle fluxes measured by low-altitude spacecraft \cite{Sergeev11, Sergeev12:IB, Dubyagin13}. In this study we investigate patterns of scattering inducing such precipitations in the current sheet embedding dipolarization front. This magnetic field configuration contains two scattering regimes: the equatorial scattering in the pre-front current sheet (that has been discussed and described in details in Refs. \cite{Birmingham84, Delcourt94:scattering, Delcourt96:choas, Shustov15}) and off-equatorial scattering at the strong $dB_z/dx$ gradient of the front. The second scattering regime is quite unusual, because it operates on the boundary of weak $B_{-z}$ and strong $B_{+z}$ fields. Within the classical models of the scattering in the current sheet, such $B_z$ increase would be interpreted as a strong reduction of precipitation \cite{Sergeev18:grl}. We demonstrate, however, that due to strong $dB_z/dx$ there are two off-equatorial regions of scattering instead of the single equatorial region. The off-equatorial $R_c$ minima will enhance scattering and may form double-peak pattern of precipitations as seen on the low-altitude spacecraft\cite{Sergeev18:grl,Dubyagin21}. Therefore, the interpretation of measurements of such double-peak patterns might be reinvestigated with including direct equatorial measurements of the current sheet configuration.

There is a problem of a model construction for precipitation patters due to scattering in the current sheet with interchanging regions of dominant $dB_x/dz$ and $dB_z/dx$ gradients. If the model of scattering in the current sheet can be parametrized by a single parameter $\kappa$, any models of scattering in the current sheet with the dipolarization front should include at least two parameters to characterize contributions of $dB_x/dz$ gradient (the equatorial $\kappa$) and $dB_z/dx$ gradient (the off-equatorial $R_c$ minimum). However, within such two parametrical model any interpretation of precipitation patterns derived from low-altitude spacecraft observations would be ambiguous, because enhancement/weakness of precipitation may be explained by two independent parameters, i.e. there are two scenarios describing any precipitation variations.

Therefore, probing of the current sheet configuration from precipitation measurements at low-altitude spacecraft should somehow distinguish between equatorial and off-equatorial scatterings. One of possible solutions is consideration of the time-scale of such scattering, i.e. energetic particle bounce period scale. The current sheet scattering is very stable and can operate for a sufficiently long time to probe it with precipitations of plasma sheet ions (e.g., the most widespread type of analysis of low-altitude precipitations is based on processing of $> 30$ keV ion fluxes measured by Polar Operational Environmental Satellites\cite{Sergeev12:IB, Dubyagin13, Sergeev15, Dubyagin18}). Scattering of such ions on $dB_z/dx$ would mix spatial and temporal effects due to comparable time-scales of dipolarization front motion/evolution and bounce oscillations of ions. This problem, however, is not actual for energetic electrons that probe instantaneous magnetic field configuration much faster than it may evolve \cite{Yahnin97}. In absence of accurate and energy/pitch-angle resolved electron measurements on existing low-altitude spacecraft such probing of the magnetotail configuration with electron measurements was not very widespread, but this approach should be more available with new low-altitude spacecraft missions \cite{Angelopoulos20:elfin, Johnson20:FIREBIRD}

In conclusion, we have investigated scattering of energetic particles in the current sheet configuration with the embedded dipolarization front. Main findings of this study are:
\begin{itemize}
\item For a weak $dB_z/dx$ gradient ($L_x/L_z \geq 1$) the magnetic field $B_z$ increase suppresses scattering as expected for the quasi-1D current sheet model.
\item For a strong $dB_z/dx$ gradient ($L_x/L_z < 0.2$) the magnetic field lines at the front have off-equatorial curvature radius minima. Energetic particle scattering at these minima changes significantly the scattering pattern: there is no monotonic dependence of the scattering efficiency on $B_z$ anymore.
\item Off-equatorial scattering forms local peaks/minima of the scattering efficiency at the front. Such peaks/minima is seen  for a wide pitch-angle range including field-aligned particles.
\end{itemize}

\section*{Acknowledgments}
The work of A.S.L. was supported by the Russian Scientific Foundation, project 17-72-20134. X.-J.Z. acknowledges support from NASA grants 80NSSC20K1270.

\section*{Data Availability}
This is theoretical study, and all figures are plotted using numerical solutions of equations provided with the paper. The data used for figures and findings in this study are available from the corresponding author upon reasonable request.

\bibliographystyle{unsrtnat}
%\bibliography{full}

\end{document}